\documentclass{elsart}
\usepackage{graphicx}

\begin{document}

\begin{frontmatter}

\title{A high power liquid hydrogen target for the Mainz A4 parity violation experiment}

          \author{I. Altarev\corauthref{asso}}
    \address{Technische Universit{\"a}t M{\"u}nchen, D-85748 M{\"u}nchen, Germany}
          \author{E.~Schilling     },
          \author{S.~Baunack       },
          \author{L.~Capozza       },
          \author{J.~Diefenbach    },
          \author{K.~Grimm         },
          \author{T.~Hammel        },
          \author{D.~von Harrach   },
          \author{Y.~Imai          },
          \author{E.~M.~Kabu{\ss}  },
          \author{R.~Kothe         },
          \author{J.~H.~Lee        },
          \author{A.~Lopes Ginja   },
          \author{F.~E.~Maas\corauthref{cor}},
          \author{A.~Sanchez~Lorente},
          \author{G.~Stephan       },
          \author{C.~Weinrich      }
    \corauth[asso]{Associated Member of the St.~Petersburg Nuclear Physics Institute, Russia}
    \corauth[cor]{Corresponding author. E-Mail: maas@kph.uni-mainz.de}
    \address{ Institut f\"ur Kernphysik, Universit\"at Mainz, D-55099, Mainz, Germany, }

%          \author{V. Lobashev    }
%    \address{Institute for Nuclear Research of RAS, Moscow, Russia}

\begin{abstract}
We present a new powerful liquid hydrogen target developed for the
precise study of parity violating electron scattering on hydrogen
and deuterium. This target has been designed to have minimal
target density fluctuations under the heat load of a 20$\mu$A CW
854.3~MeV electron beam without rastering the electron beam.
The target cell has a wide aperture for scattered electrons and is axially
symmetric around the beam axis. The construction is optimized to
intensify heat exchange by a transverse turbulent mixing in the
hydrogen stream, which is directed along the electron beam. The
target is constructed as a closed loop circulating system cooled
by a helium refrigerator. It is operated by a tangential
mechanical pump with an optional natural convection mode. The
cooling system supports up to 250 watts of the beam heating
removal. Deeply subcooled liquid hydrogen is used for keeping the
in-beam temperature below the boiling point. The target density
fluctuations are found to be at the level 10$^{-3}$ at a beam
current of 20 $\mu$A.
\end{abstract}


\end{frontmatter}

\section{Introduction}
Cryogenic liquid hydrogen targets have become common
experimental equipment in and particle physics experiments.
The most powerful new targets have been requested for
electron scattering parity violation experiments,
which are able to register the small effects of weak interaction at the level
10$^{-6}$. In these targets heating can reach a few hundred watts,
but target density fluctuations should not become larger than
10$^{-3}$ to reach high statistical accuracy of
measurements. One source of target density fluctuations is the
heating of the liquid hydrogen by the energy deposit of
the electron beam. Overheating can cause boiling. Therefore, the task of
in-beam heat removal is to be solved.
Recently, Experiments at the MIT-Bates accelerator \cite{sample:target,sample:physics},
at the TJNAF \cite{happex:physics}, and at SLAC \cite{e158:physics}
have reported on their powerful targets.
In this paper we present the target design and realization for the
A4-experiment at the MAMI accelerator in Mainz
\cite{maas:pv:2004,maas:2photon:2005,maas:pv:2005,maas:a4cal:2003,maas:a4cal:2002}.

The aim of the A4-experiment is a precise measurement of the
parity violating asymmetry in the
elastic scattering of right and left handed electrons  on an
unpolarized proton target. The expected
value of the asymmetry is on the order of 10$^{-6}$.
The experiment is carried out at the MAMI facility in Mainz  with
a 80\% polarized 854.3 Mev electron beam. With modern GaAs photocathodes
high currents of polarized electron beam combined with online intensity
and polarization monitoring have been realized \cite{polsource:aulenbacher:97}.
A large calorimeter with 1022 crystals
to handle a total of 100~MHz counting rate has been developed. An
innovative feature of this detector is a use of new, high
resolution and very fast PbF$_{2}$ Cerenkov radiators \cite{pbf2:achenbach:01}.
The 10~cm liquid hydrogen target provides a luminosity of
$5 \times 10^{37}~$cm$^{-2}$s$^{-1}$ at 20~$\mu$A electron beam current.
This corresponds to about 100 watts of heat absorbed in the
liquid hydrogen and in the aluminum windows.
The target cell is included in a closed loop with
a hydrogen-helium heat exchanger, where the liquid hydrogen
circulation is forced by a mechanical tangential pump. This is a common
scheme for the high power
cryogenic system in which a high mass flow and a high heat
transfer rate can be achieved.
We outline the characteristic features of this target as follows.
\begin{itemize}

\item[-] The target has a $140^\circ$ wide aperture for scattered
electrons in forward (optionally in backward) direction. It is
axially symmetric with respect to the incident beam axis.

\item[-] The hydrogen moves along the beam. Therefore a special
flow geometry was applied to intensify a turbulent transverse mixing in
the hydrogen stream. The turbulence causes a mass exchange and a
heat removal from the heated region transverse to the beam
direction.

\item[-]  The helium cooling system was designed in a way to maintain
deeply subcooled (down to freezing point) liquid hydrogen in
order to enlarge the head room for the beam heating without boiling.

\item[-]  A powerful cryogenic pump was used. It produces up to
0.3 bar overpressure and can provide a  mass flow of several hundred
grams per second of liquid hydrogen. This would cover the needs of a
few kilowatt cryogenic system, provided the corresponding helium
refrigerator power is provided.

\item[-]  The circulation contour was designed in a way to apply the
largest available pressure decline very locally at the target cell
right to the region of beam heating. To this end a low flow resistance
hydrogen-helium heat exchanger was developed.

\item[-]  The target works well even without running the mechanical pump
up to 120 watts heat load since an intensive natural convection driven
circulation takes place in the loop. This regime is useful for applications
when the density
fluctuations do not play an important role.

\item[-]  The target gas system is simple and reliable. It contains
only 5 m$^{3}$  of hydrogen at 2.2 bar of absolute pressure. All
safety requirements, conventional for the accelerator environment,
are fulfilled and the target it inherently safe in case of a sudden
loss of cooling.

\end{itemize}
In the following sections we explain our design approach, give a
description of the target and present the results obtained.

\section{Local beam heating and luminosity fluctuations}
\label{chap:beamheating} In this section we aim
at an estimate of the luminosity fluctuations caused by
temperature variations in the hydrogen stream. We try to reveal
the essential parameters for the effective suppression of the
fluctuations considering a primitive model of turbulence, since
the exact treatment of fluid dynamics in
the turbulent regime is difficult and would require detailed
numerical simulations.
The numerical estimates are based on an electron beam
with 854.3~MeV energy and 35~$\mu$A. It deposits a heating power
of 160~W into the liquid hydrogen target. We assume a hydrogen
flow in a tube coaxial with the electron beam. The diameter of the
tube is 1.2~cm.

\subsection{Estimation of local beam heating}
Following Prandtl's concept of turbulent
mixing length \cite{handbook:turbulence:1977} we have the space scale of turbulent local fluctuations as
\begin{eqnarray}
   l_t &=& \kappa \; y
\end{eqnarray}
where $\kappa=0.4$ is an empiric constant and
$y$ is the distance from a parallel wall to the flow direction.
This corresponds to half of a tube diameter $D$. In our case
the resulting turbulent mixing length $l_t=2.4$ mm, which is of order or larger than
the possible electron beam diameter $d$ in mm.
We can also apply the
concept of relaxation time $t_t$ of the turbulent transfer,
that defines the time scale of turbulent fluctuations:
\begin{eqnarray}
   t_t=l_t/v^{\ast}
\end{eqnarray}
with the dynamic
velocity $v^{\ast} = v\sqrt{\zeta /2}$, the friction coefficient
$\zeta=0.316/\sqrt[4]{Re}$, and the Reynolds number $Re$.
In our case the relaxation time is in the region
of several milliseconds. For example, for hydrogen of T=16~K and
$v=2$~m/s the relaxation time results in $t_t=12.5$~ms.
In this simple model any existing local
fluctuation, for example temperature fluctuation $\Delta T_f$,
will exponentially dissipate with time $t$ as
\begin{eqnarray}
\Delta T(t)&=&\Delta T_f \; e^{-(t/t_t)}. \label{eq:deltat:expo}
\end{eqnarray}
The heating due to the electron beam occurs through Bethe-Bloch
energy deposit at a constant specific volume energy deposition
rate $q_v$. Fluctuations of temperature arises from cold
fluid entering the beam and dissipate as the heated mass is replaced by
cold mass from the region outside the beam. For a static case,
the temperature rise due to the beam heating
is
\begin{eqnarray}
\Delta T_f(t)&=&\frac{q_v}{C_p\;\rho} \; t,
\end{eqnarray}
where $C_p$ denotes the specific liquid hydrogen
heat capacity and $\rho$ the hydrogen density. Combining this with
Eq.~\ref{eq:deltat:expo}, we get a time structure of the
temperature fluctuations of
\begin{figure}[b]
   \begin{center}
   \includegraphics[width=0.6 \textwidth]{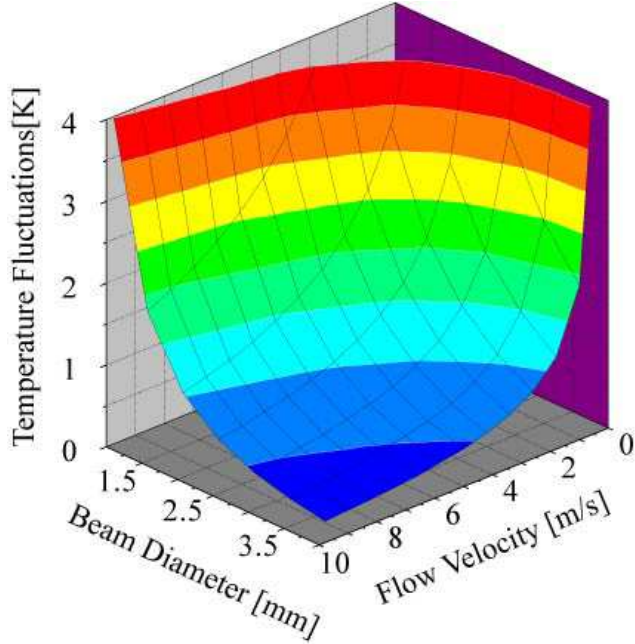}
   \end{center}
   \caption{Beam heating for 12~mm tube.}
  \label{DeltaT}
\end{figure}
\begin{eqnarray}
\Delta T(t)=\frac{q_v}{C_p\;\rho} \; t \; e^{-(t/t_t)}.
\end{eqnarray}
This function has a maximum at $t=t_t$. This gives the amplitude of
temperature fluctuations as
\begin{eqnarray}
\Delta T_{\mathrm{max}}= \frac{q_v}{C_p\;\rho} \; t_t \; e^{-1},
\end{eqnarray}
which drops with the beam size (due to $q_v$) and with flow velocity as shown at
Fig.~\ref{DeltaT}. An efficient heat transfer from the beam area
takes place when both $d>2$ mm and  $v>2$ m/s.
\subsection{Luminosity fluctuations without boiling}
Even without boiling, luminosity
fluctuations $\Delta \mathcal{L}/\mathcal{L}$ can originate from target density
fluctuations $\Delta \rho / \rho$ arising from temperature
fluctuations. The scale of these fluctuations is
given by the value of $\Delta T_{\mathrm{max}}$. An accomplishable
value, is $\Delta T_{\mathrm{max}}\approx 1$~K, as one can see from
Fig.~\ref{DeltaT}. The corresponding relative luminosity
variation amounts to
\begin{eqnarray}
(\frac{\Delta \mathcal{L}}{\mathcal{L}})_{\mathrm{single}} &=& \frac{\Delta
\rho}{\rho} \approx  0.014
\end{eqnarray}
for 16~K hydrogen.
This estimate is for a single turbulent
fluctuation with size of order $l_t$. In order to get an estimate
of the overall luminosity fluctuations over the whole target
length, we average the single fluctuations over the
the beam volume containing a number $n$ of single fluctuations
contributing to beam scattering during the integration time window
$t_i = 20$~ms. Since $l_t>d$,
the number of simultaneous fluctuations penetrated by the beam is
$l/l_t$, where $l$ is the target length of 10~cm. This value has
to be integrated over $t_i$, yielding
\begin{eqnarray}
n &=& \frac{l}{l_t} \frac{t_i}{t_t}.
\end{eqnarray}
With the assumptions as
stated above, the total luminosity fluctuations result in
\begin{eqnarray}
(\frac{\Delta L}{L})_{\mathrm{total}} &=& (\frac{\Delta
L}{L})_{\mathrm{single}} / \sqrt{n} \approx 1.7 \cdot 10^{-3}.
\end{eqnarray}
This estimate for the luminosity fluctuations is of the same order
as required by the experiment. It can be much larger if local boiling
takes place.

\subsection{Boiling at the aluminum windows}

The electron beam deposits energy not only in the hydrogen volume,
but also in the thin aluminum entrance and exit windows. This
heating amount to 1.9~W for  0.1~mm of aluminum thickness which
corresponds to a specific heat flow from the beam spot at the
aluminum window of $q_f=60\,W/$cm$^2$ for beam
diameter of $d=2$~mm. A more detailed analysis shows that the heat is
not transferred directly to the liquid, but is spread over the
aluminum window due to the aluminum heat conductivity.
Nevertheless, the specific heat flow from the aluminum to the
liquid hydrogen does not get below $q_f \approx$ 5~W/cm$^2$ to 10~W/cm$^2$.
This exceeds the first boiling crisis value ($\sim 1.5$~W/cm$^2$) \cite{handbook:cryo:1970}
and causes unavoidable boiling around the beam spot in the
hydrogen volume right at the aluminum window. It should be noted
that at such a large value of $q_f$ the boiling regime is not
stable in time. The wall temperature is also unstable and its
average value is determined mainly by the heat spreading over the
wall trough the bulk aluminum and can, in our case, reach a few
ten degrees. This process does not depend much on beam size or on
aluminum wall thickness. With increasing flow velocity its
contribution to the luminosity fluctuations will be proportionally
less.
\subsection{Design criteria summary}
We summarize the target efficiency parameters as they follow from
our estimations which is a delicate problem down to the level of
fluctuations of $10^{-3}$. We need a beam
size of a few millimeters in diameter as an alternative to the
method of beam rastering. The temperature of liquid hydrogen
should be close to the freezing point in order to give several
degrees of headroom to the boiling temperature. The cross section
of the tube (nozzle), which guides the hydrogen stream along the beam,
should be minimized. This minimum has to provide the necessary
mass flow (a mass flow of 20~g/s gives a
1 degree rise of the bulk hydrogen
temperature at 160~W heating power). The flow velocity
has to be maximized. A limiting value is
defined by the pressure drop in the nozzle.
They are evaluated as a local pressure decline
\begin{eqnarray}
\Delta P=\xi \frac{\rho v^2}{2}
\label{forc}
\end{eqnarray}
with the coefficient $\xi \sim 2$.
This is a central item for the
design of the heat exchanger cooling system since it requires a
careful optimization with respect to the flow
resistance, which has to be small compared with
the value given by formula \ref{forc} for the remaining parts in the hydrogen system.
\section{Experimental setup}
\begin{figure}[hbt]
   \begin{center}
   \includegraphics[width=0.8\textwidth]{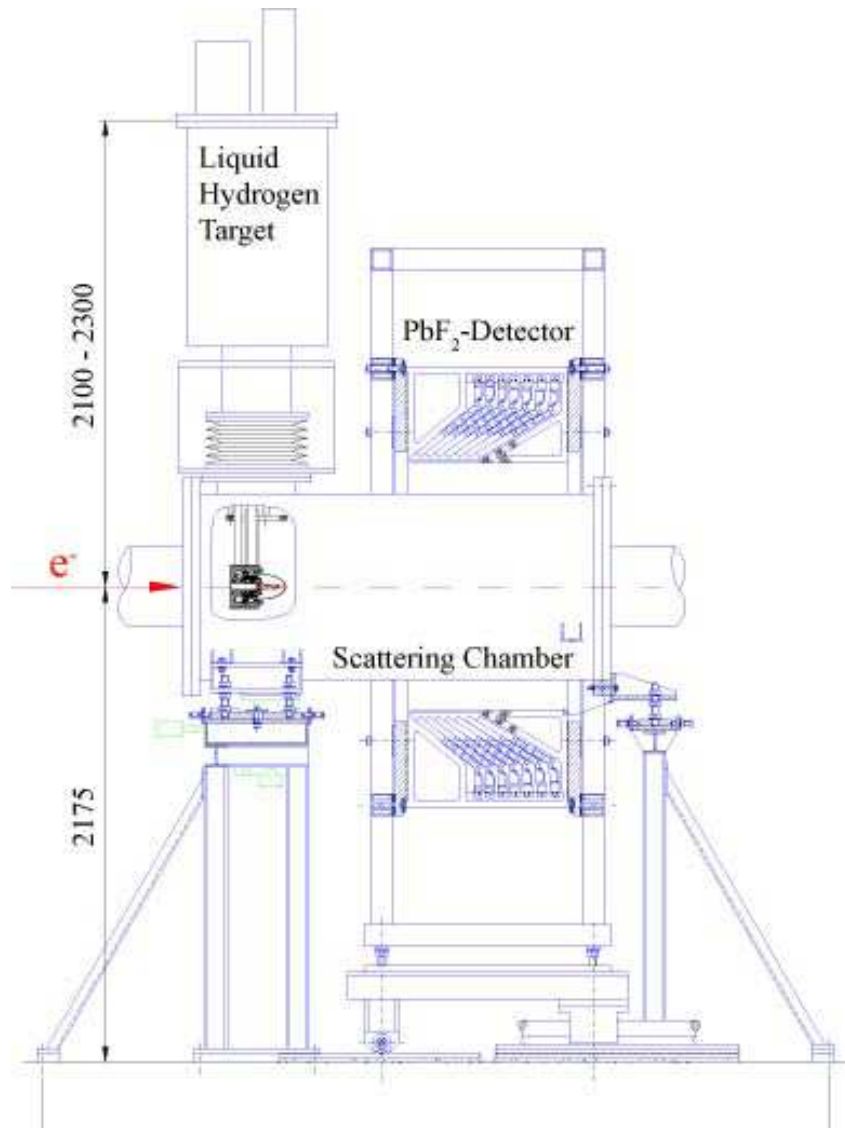}
   \end{center}
   \caption{The experimental setup of the Mainz A4 parity experiment.}
  \label{exp}
\end{figure}
Fig.~\ref{exp} shows the experimental setup. The electron beam
enters from the left and hits the high power liquid hydrogen
target in the vacuum vessel of the scattering chamber. The
scattering chamber is surrounded by the large acceptance (0.64~sr)
multichannel PbF$_2$-detector that registers electrons in the
electron scattering angle range of $30^{\circ} < \theta_e <
40^{\circ}$ \cite{maas:a4cal:2002}. The luminosity is measured
under small scattering angles of $4.4^\circ < \theta_e <
10^\circ$. The target cryostat is placed on an
elevator system. The large elevator bellow allows a 20~cm
expansion along the vertical axis in order to move the target in
and out of the beam line. An adjustable incline of the axis under
any azimuthal angle is possible, so that the target cell can be
aligned.
\section{The target cell}
The axial symmetry of the detector which is required by the physics
of the A4 parity experiment is also reflected in the liquid hydrogen
target cell design.
\begin{figure}[hbt]
   \begin{center}
   \includegraphics[width=0.8\textwidth]{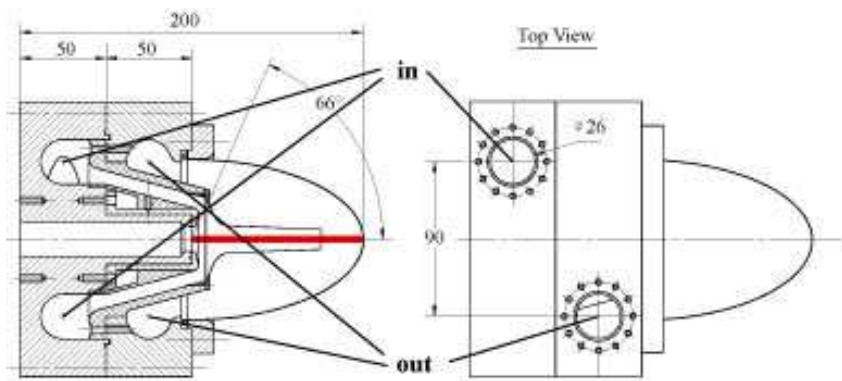}
   \end{center}
   \caption{Target cell design of the liquid hydrogen target. Left: view
   from the side. Right: view from top. }
  \label{cell}
\end{figure}
Fig.~\ref{cell} shows the target cell design. The target consists of
aluminum parts joined with an indium sealing, so that the
target can be reassembled many times. The target thickness is
limited to 10 cm for forward scattering experiments by the variation
of the scattered electron energy.
The liquid hydrogen fluid enters the toroid shaped collector. The
hydrogen flow is directed into a nozzle of 12 mm diameter along the
beam path, where the maximum flow velocity and the maximum degree of turbulence
is reached. The fluid leaves the target cell through the exit collector.
The aluminum entrance window has a thickness of 75 $\mu$m. The large parabolic
aluminum cap which contains the hydrogen has a regular wall thickness of
about 250 $\mu$m, but at the place of the beam exit it is thinned to 100 $\mu$m.
The nozzle walls are 200 $\mu$m thick, so that a cone of up to 66$^\circ$
has a low thickness of material in the way of the scattered electrons.
This construction has been tested up to 6~bar pressure in the cell against vacuum.
The target cell is fixed through a thermally isolating bridge to
the bottom cryostat flange in order to minimize mechanical vibrations and
temperature variations.

The heating by the electron beam is concentrated in the small
volume of the electron beam path. As the flow is directed along the
beam axis, the only efficient mechanism of heat removal from this
overheated volume is the turbulent transverse mixing, which is more
efficient at higher velocity. This mechanism was considered in section
\ref{chap:beamheating}.
An even smaller nozzle diameter would enhance the degree of
turbulence to such an extend, that the mass flow along the beam
would drop too much. The nozzle cross section must be large
enough to avoid additional background arising from scattered electrons at
small scattering angles.
\section{The target cooling system}
\subsection{Cooling principle}
\begin{figure}[h]
   \begin{center}
   \includegraphics[width=0.8\textwidth]{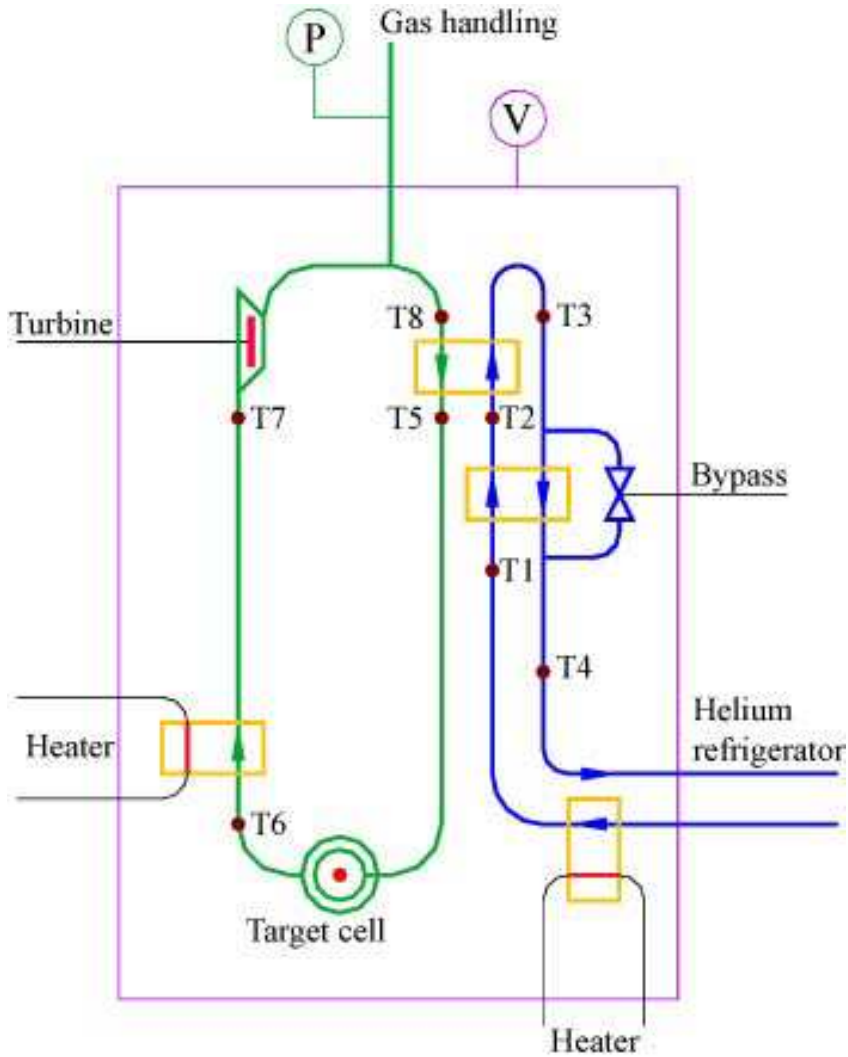}
   \end{center}
   \caption{Schematic of the target cooling system.}
  \label{coolpr}
\end{figure}

A schematic of the target cooling system is shown
in Fig.~\ref{coolpr}.
The system consists of a closed  liquid
hydrogen circulation loop and a cold helium gas cooling loop
including a 400 watt Linde helium refrigerator. For the heat
transfer from hydrogen to helium we use a counterflow heat
exchanger, placed at the top of the hydrogen loop. The hydrogen
target cell is located at the lowest point in the loop. The liquid
hydrogen circulates clockwise driven by a tangential pump from the
target through the pump to the heat exchanger top. After cooling
in the heat exchanger it flows again to the target. In order to
stabilize the total heat load of the refrigerator a temperature
controlled electrical heater is added in the loop right behind the
target. This provides a stable operation at variable electron beam
currents and provides a heat load even in case the beam is turned
off. The loop hight is about 2 m. For safety
reasons, the hydrogen loop is always connected to an external gas tank of 5
m$^3$ at room temperature. The hydrogen system in total
contains, including the gas tank, an amount of hydrogen corresponding to 2.2 bar of
absolute pressure at room temperature. When the helium temperature
gets below the hydrogen boiling point, the heat exchanger works
as a liquifier until the loop is completely filled with liquid
hydrogen. The loop volume of liquid hydrogen is about 5 l,
therefore the pressure in the system drops to 1.2 bar. When the
liquid covers the heat exchanger completely, it will be cooled
down further to helium temperature since the condensation process
stops. When there is no cooling the liquid is naturally evaporated
filling the buffer gas tank back to the 2.2 bar.
There is an additional
electrical heater in the helium loop, in order to maintain the
temperature level of incoming helium at an appropriate level.
In the helium-helium heat exchanger incoming cold
helium gas is heated by warm Helium, coming from the
helium-hydrogen heat exchanger.
The additional helium-helium heat exchanger,
controlled by a bypass valve, allows further regulation
of the temperature at the working point without
changing the load on the helium refrigerator. This is
especially useful when deuterium is used instead of hydrogen.
The system has eight temperature sensors. It is
controlled by a computer with an interface card having analogue to
digital converters (ADC) and digital to analogue converters (DAC).
Two modes of operation are possible. A soft mode, where the
control system regulates only the power of the helium heater to
keep the incoming helium temperature T$_2$ right above the
hydrogen freezing point ($\sim$15 K). It is convenient for
automatic target cool-down and for operation in natural convection
mode. When operating with the electron beam on the target, the
control system regulates both the power of the hydrogen heater
(T$_8$ is kept stable) and the incoming helium temperature T$_2$.
\subsection{Natural convection}
The hydrogen loop has been optimized to provide the conditions for
a hydrogen circulation based on natural convection driven by the
density differences of cool and hot hydrogen liquid. For many
target applications this regime is quite sufficient. It is known
\cite{cns}, that optimized natural convection can provide
significant heat removal, up to a few kilowatts of power. For the
optimization high flow resistance parts in the loop have been
removed such as pipe bends, sharp changes of cross sections and so
on. As a result of this approach about 2/3 of the total pressure
loss along the whole loop comes from the target cell. For natural
convection the difference of static pressure $\Delta P_{st}$
between cold (on the right side) and warm (on the left) loop
branch should be compensated by the dynamic pressure loss $\Delta
P$ over the whole circulation loop. The static pressure difference
can be calculated from the density difference $\Delta \rho$
\begin{eqnarray}
\Delta P_{st} = g \; \Delta \rho \; L
\end{eqnarray}
where $g = 9.81$~m/s$^2$ and $L \approx 2$~m is the loop height.
The maximum of  $\Delta \rho$ is achieved when the hydrogen
temperature in the cold branch is close to freezing temperature
and the warm branch is close to boiling temperature. From this
follows that $\Delta P_{st}$ is of order
1~mbar and 2/3 of that have to be applied in the nozzle. This
corresponds to a local pressure loss caused mainly by narrowing
of the stream before and its expansion after the nozzle. It is proportional to
the square of the flow velocity according to formula \ref{forc}.
With a 12~mm nozzle diameter we  achieve a flow velocity of
$v=0.9$~m/s and a hydrogen mass flow of $G \approx 7.5$~g/s. These
values provide a stable operation of the target in natural
convection mode, but they are not sufficient to suppress luminosity
fluctuations down to the required level of $10^{-3}$. Operation
with a mechanical tangential pump gives a significant improvement
since the possible pressure drop increases due to a pump
about 300 times reaching 0.3~bar.
A sufficiently high velocity of 15~m/s, as estimated from
\begin{eqnarray}
\frac{v_{pump}}{v_{nat}}\sim \sqrt{\frac{\Delta P_{pump}}{\Delta
P_{st}}},
\end{eqnarray}
and a mass flow of 130~g/s is possible in this case.
\subsection{Construction design of the hydrogen loop system}
\begin{figure}[h]
   \begin{center}
   \includegraphics[height=0.7\textheight]{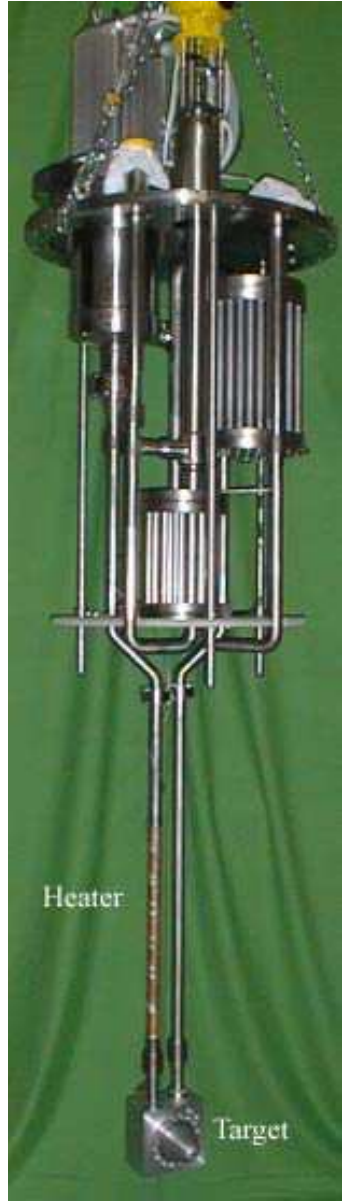}
   \end{center}
   \caption{Photograph of the hydrogen circulation and cooling loop system.}
  \label{lht}
\end{figure}

The liquid hydrogen loop and target cooling system is incorporated
in one cryostat with all inputs going through a top flange.
Fig.~\ref{lht} shows a photograph of the inner part
of the cooling system without the super isolation. The photograph
corresponds to the elements in the diagram of Fig.~\ref{coolpr}.
The target cell at the bottom of the picture is connected on the left
side to the hydrogen turbine and on the right side to the main hydrogen-helium heat
exchanger located right under the top flange. This heat exchanger
consists of 78 tube-in-tube elements with total heat transfer
surface about $1.4$~m$^{2}$. The helium-helium heat exchanger
in the center is smaller. It contains 32 short tube-in-tube
elements. The total height of the loop is about 2~m, the
top flange diameter 0.6~m. The hydrogen target cell is made from aluminum,
all other parts are made from stainless steel.
Joints are made using conflat flanges with copper gaskets unless welded.
The system is fixed to the top flange by the cold valve body, attached
to the helium-helium heat exchanger top collector. All thermal length changes
are compensated by  bellows.
As a hydrogen pump (on the left of the main heat exchanger) we use
the Barber-Nichols tangential pump BNHeP-15-000 giving a closed
pressure difference of 0.3~bar. It has originally been designed
for high flow rate (up to 600 g/s at 100 Hz) transfer of liquid
helium. We run the pump at a rotation speed of 70~Hz. We estimate
the heating of the hydrogen by the pump as $\sim 20$~W
corresponding to about half of the total parasitic heat load of
the system. The additional liquid hydrogen volume required by the
tangential pump corresponds to about 1~l. The turbine can be
substituted by a short piece of tube if only the natural
convection mode is needed. A 400 watt Linde helium refrigerator is
available for target cooling. It produces a flow of cold helium
gas at a temperature of 7-11~K. Helium transfer lines of 18 m
length deliver the coolant to the experimental hall. The helium
mass flow is about 6-8~g/s. Though the nominal refrigerator power
is rather high, it cannot be utilized for the hydrogen system
because of the poor mass flow. That puts  currently limit on the
cooling power of below 250~W, though the cooling system design
allows to use a helium mass flow of up to 30 g/s providing a
possible cooling power of up to 1000~W. The heat exchanger design
allows freezing of the liquid in it without safety hazard. This is
important as the
input helium temperature can be safely reduced to 12 K
and below.
%The only
%care should be taken to prevent a blocking of the circulation,
%that never happens while the turbine is running and heater is on.
\section{The gas system}
\begin{figure}[h]
   \begin{center}
   \includegraphics[width=\textwidth]{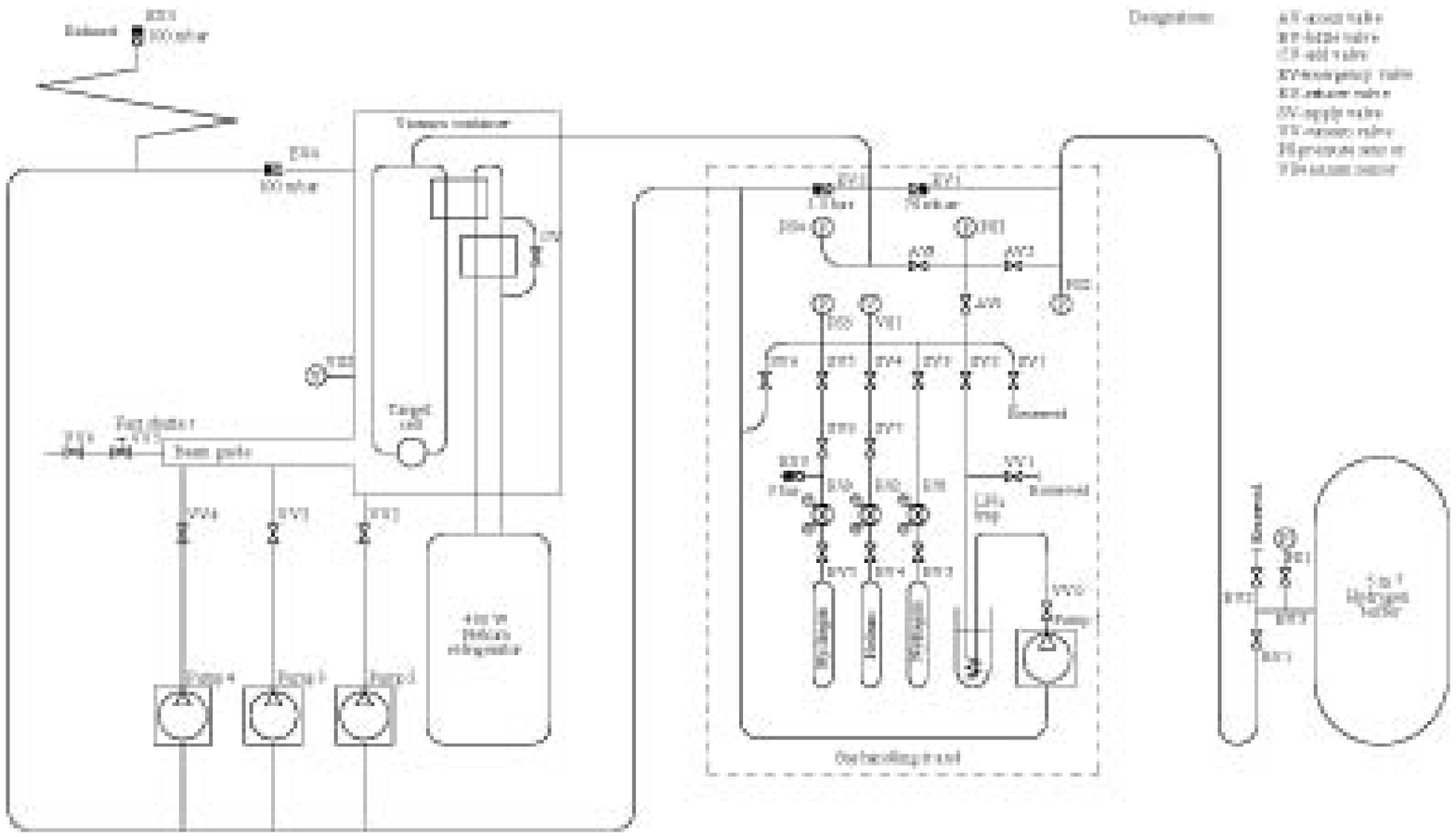}
   \end{center}
   \caption{Target operation scheme.}
  \label{gas}
\end{figure}

Fig.~\ref{gas} presents an operational flow
diagram of the complete hydrogen target system.
The cryogenic part is located in the vacuum container
formed by the target cryostat united with the scattering vacuum
chamber and the attached beam guide. The cold hydrogen loop is
connected to the 5~m$^3$ hydrogen gas buffer tank through the gas
handling stand. For the standard operation of the cold target the
valves AV2, AV3 and BV1 have to be open providing a free transfer
of hydrogen between the cold part and the buffer tank.
This connection is ensured by a one-way release
valve EV1. Access to the hydrogen system with vacuum pump and
various gases for its preparation and filling is achieved from an
access valve AV1. The safety valve EV2 prevents
overfilling of the system by venting the hydrogen to the exhaust
line on the building roof. The vent tube is kept filled with
nitrogen gas that is trapped at its far end by 100~mbar release
valve EV5 in order to keep the line always free of oxygen from
air.
In the case of a target crash the beam guide will be closed
by the fast shutter valve VV5 preventing hydrogen from entering
the high power microwave system of the accelerator. In the case of
hydrogen escape to the experimental hall two hydrogen sensors
control a special venting procedure of the experimental hall.
\section{Target control}
The target control system includes the readout of all the
operation parameters (16 input channels), the control of two
feedback loops for regulating the helium and the hydrogen heater
power, and the control of the cold helium bypass valve. The scheme
is realized with a computer interface card DAP
2400e/5 board of Microstar Laboratories. The board is controlled
by DIGIS software on a single PC. The helium refrigerator has an
independent automatic control and readout. The temperature sensors
(see Fig.~\ref{coolpr}) are standard silicon diode sensors, except
$T_6$, which is a carbon glass resistor. This type of sensor has a
high radiation resistance, which is necessary for the a
temperature sensor located near to a target cell.

\section{Test results}

\subsection{Tests with the electrical heater in the hydrogen circuit}
In order to commission the hydrogen target cooling system we
operated the system after liquifying the hydrogen gas without the
electron beam, only using the electrical heater in
the hydrogen loop. All target cooling system parameters like the temperatures,
the heating power of both heaters, the valve positions etc.
(as displayed in Fig.~\ref{coolpr}) have been measured
during the tests. The efficiency of the hydrogen heater is 100~\% i.e. all the
electrical power is transferred to the hydrogen liquid. It simulates the global
heat load of the electron beam.
\begin{figure}[htbp]
   \begin{center}
   \includegraphics[width=\textwidth]{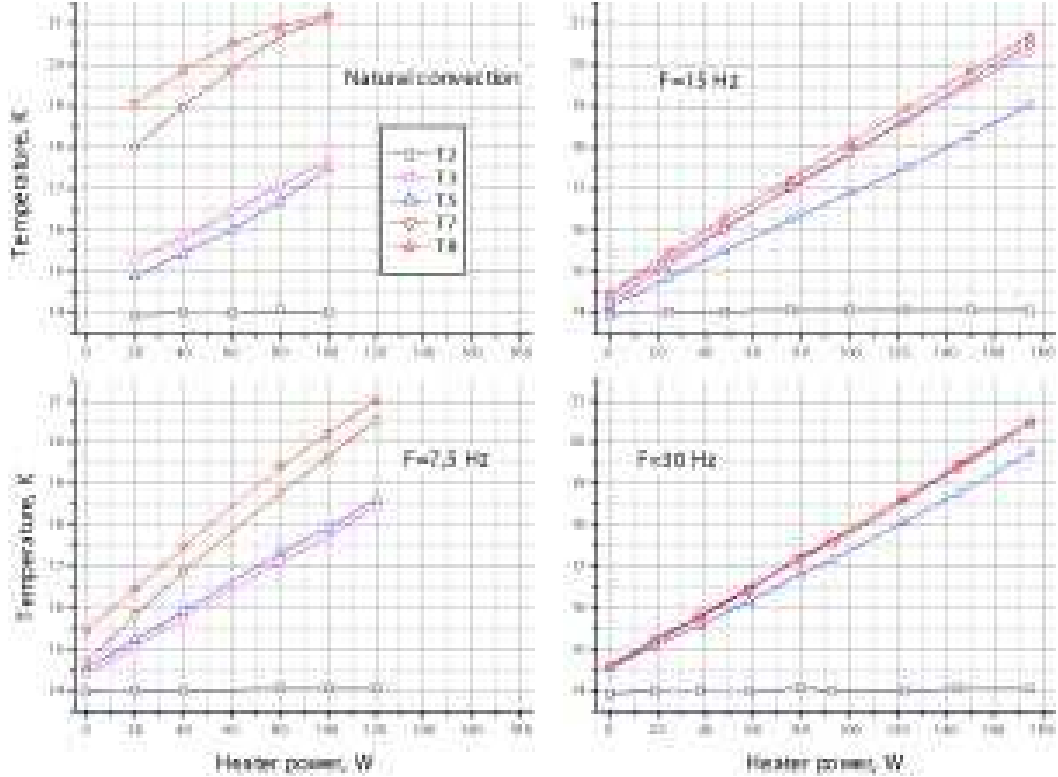}
   \end{center}
   \caption{Target performance.}
  \label{fig:nat}
\end{figure}
For a high heat load of 250~W the highest temperature in the system T8
behind the target rises 19.7~K which is still about 1.3~K below the boiling point of
hydrogen. The target entrance temperature T5 for this high heat
load is 18~K corresponding to 3~K subcooled hydrogen.
For working conditions corresponding to a more realistic heat load
of 100~W and the tangential pump rotation frequency to 50~Hz,
the maximum system temperature T8 settles at 17.3~K. The target entrance temperature
T5 stabilizes at 16~K corresponding to 5~K subcooled hydrogen.
These first tests show rather quantitatively, that the target can
be operated at the working conditions of the A4 experiment.

In order to determine the efficiency of the hydrogen-helium heat
exchanger we have studied the temperature variation in the system
for different tangential pump rotation speeds (corresponding to
different values of the hydrogen mass flow) as a function of the
electrical heater power. The four plots in Fig.~\ref{fig:nat} show
the results for T2, T3, T5, T7, and T8 for natural convection
(lowest mass flow rate), tangential pump speed of 7.5~Hz, 15~Hz,
and 30~Hz. T2 has been stabilized to 14~K for all four pump
speeds. T2 corresponds to the helium inlet temperature of the
hydrogen-helium counterflow heat exchanger and T3 corresponds to
the helium outlet temperature (see Fig.~\ref{coolpr}) whereas T8
corresponds to the hydrogen inlet temperature and T5 to the
hydrogen outlet temperature.
The difference $\Delta$T$_{\mathrm{He}}$=T3-T2 is proportional to
the cooling power at constant helium mass flow.
One sees that for all four mass flow values the helium cooling power
( which is proportional to $\Delta$T$_{\mathrm{He}}$)
varies linearly with the electrical heater power.
Only for the case of natural convection, the difference shows nonlinear
behavior due to the fact, that the mass flow for natural convection is a
function of the temperature. The difference of T8-T7 is proportional to the heat
deposition coming from the tangential pump.
When the heater is off, $\Delta$T$_{\mathrm{He}}\approx 0.5$ degree that corresponds
to about 15~W of parasitic heating. This parasitic heating is
sufficiently small and does not distinctly depend on the pump
speed. The outlet helium temperature T3 linearly rises as the heating
power approaches the hydrogen boiling temperature ($\sim$ 21 K) at
about 200 W. The Helium is heated by hydrogen, therefore the inlet
hydrogen temperature T8 is always higher than the outlet helium
temperature T3. The data for 30 Hz pump speed show a very
efficient heat transfer since T8$\approx$T3. The efficiency becomes
worse as the speed reduces. In case of natural convection the
difference T8-T3 arrives at about 3 degree being almost
independent from the heater power. The maximal hydrogen
temperature depends on the outlet helium temperature
which is, in its turn, defined by the helium mass flow. The
temperature T5, at which hydrogen comes into the target cell, is
limited by T8.
This test with the heater demonstrates that the target cooling
loop performs like expected from our estimates based on simple
fluid dynamics rules. The overall performance of the target
cooling loop is currently limited only by the available helium
refrigerator.
\subsection{Tests of the target cooling system with the MAMI 85~MeV electron beam.}
\subsubsection{Influence of Hydrogen Mass Flow and Temperature on the Target Density Fluctuations}
We have studied the target density fluctuations in the liquid hydrogen target
by hitting the hydrogen target cell with the MAMI 854.3~MeV electron beam
\cite{MAMI:euteneuer:94} at different target cooling system conditions.
\begin{figure}[hbtp]
   \begin{center}
   \includegraphics[height=0.5\textheight]{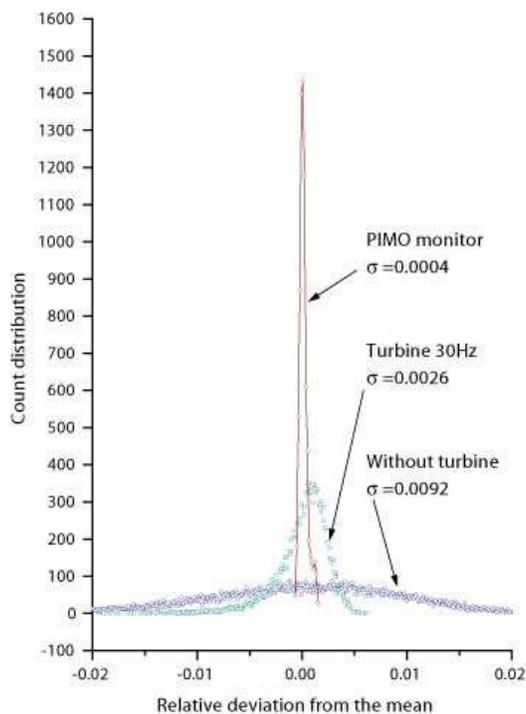}
   \end{center}
   \caption{Target density fluctuations as measured in 20~ms samples with a small
   cross section 20 $\mu$A electron beam at different pump speeds (corresponding to different
   hydrogen mass flow).}
  \label{fluc}
\end{figure}
For these tests we have measured the luminosity
$\mathcal{L}$(which is the product of electron beam current and
target density) with a system of water Cherenkov detectors
detecting electrons from M{\o}ller scattering at electron
scattering angles 4.4$^\circ$ $<$ $\theta_e$ $<$ 10$^\circ$
\cite{lumi:hammel:2005}. We have measured the electron beam
current simultaneously. During the tests, the cross section of the
electron beam has been kept small with an approximate size of
3$\times$~10$^4$~$\mu$m$^2$.
In Fig.~\ref{fluc} we show on overlay of three histograms. The
peak labeled \lq\lq Without turbine\rq\rq\ shows the distribution
of the luminosity signal for the case where the tangential pump
was not rotating and the loop was circulating in natural
convection mode with low mass flow. The measured normalized RMS
for this case amounts to $\sigma$~=~$\delta
\mathcal{L}/\mathcal{L}$~=~0.0092. The peak labeled \lq\lq PIMO
monitor\rq\rq\ shows a histogram of the beam current monitor
signal. The normalized RMS of the beam current $\mathcal{I}$ for
this measurement has been $\sigma$~=~$\delta
\mathcal{I}/\mathcal{I}$~=~0.0004 which is a factor 23 smaller
than the luminosity fluctuations, showing that the observed
luminosity fluctuations arise to a high extend from target density
fluctuations only. The peak labeled  \lq\lq Turbine 30~Hz\rq\rq\
shows the luminosity distribution for the case when the tangential
pump was running at 30~Hz giving a higher hydrogen mass flow. The
effect of the turbulent mixing in the hydrogen target cell leading
to a higher effective heat transfer can be clearly seen,
\begin{figure}[b]
   \begin{center}
   \includegraphics[width=0.6\textwidth]{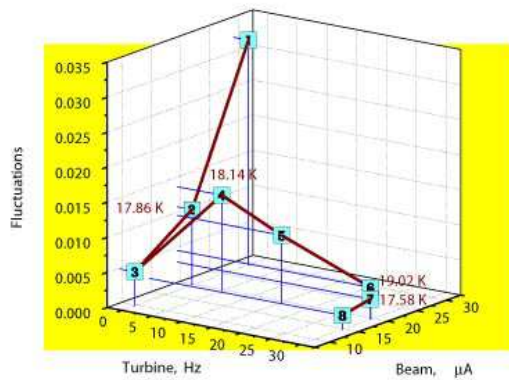}
   \end{center}
   \caption{Target density fluctuations at various conditions of tangential pump
   rotation speed and electron beam current.}
  \label{fig:fluall}
\end{figure}
since the normalized RMS of the luminosity
distribution with high mass flow is
$\sigma$~=~$\delta \mathcal{L}/\mathcal{L}$~=~0.0026 corresponding
to a factor of 3.5 decrease in the width of the target density
distribution. We have made more tests varying the
pump rotation speed and the beam current. Fig.~\ref{fig:fluall}
shows some of the results. One recognizes that the normalized RMS
of the target luminosity varies from $\sigma$~=~$\delta
\mathcal{L}/\mathcal{L}$~=~0.032 at 30~$\mu$A
\begin{figure}[t]
   \begin{center}
   \includegraphics[width=0.6\textwidth]{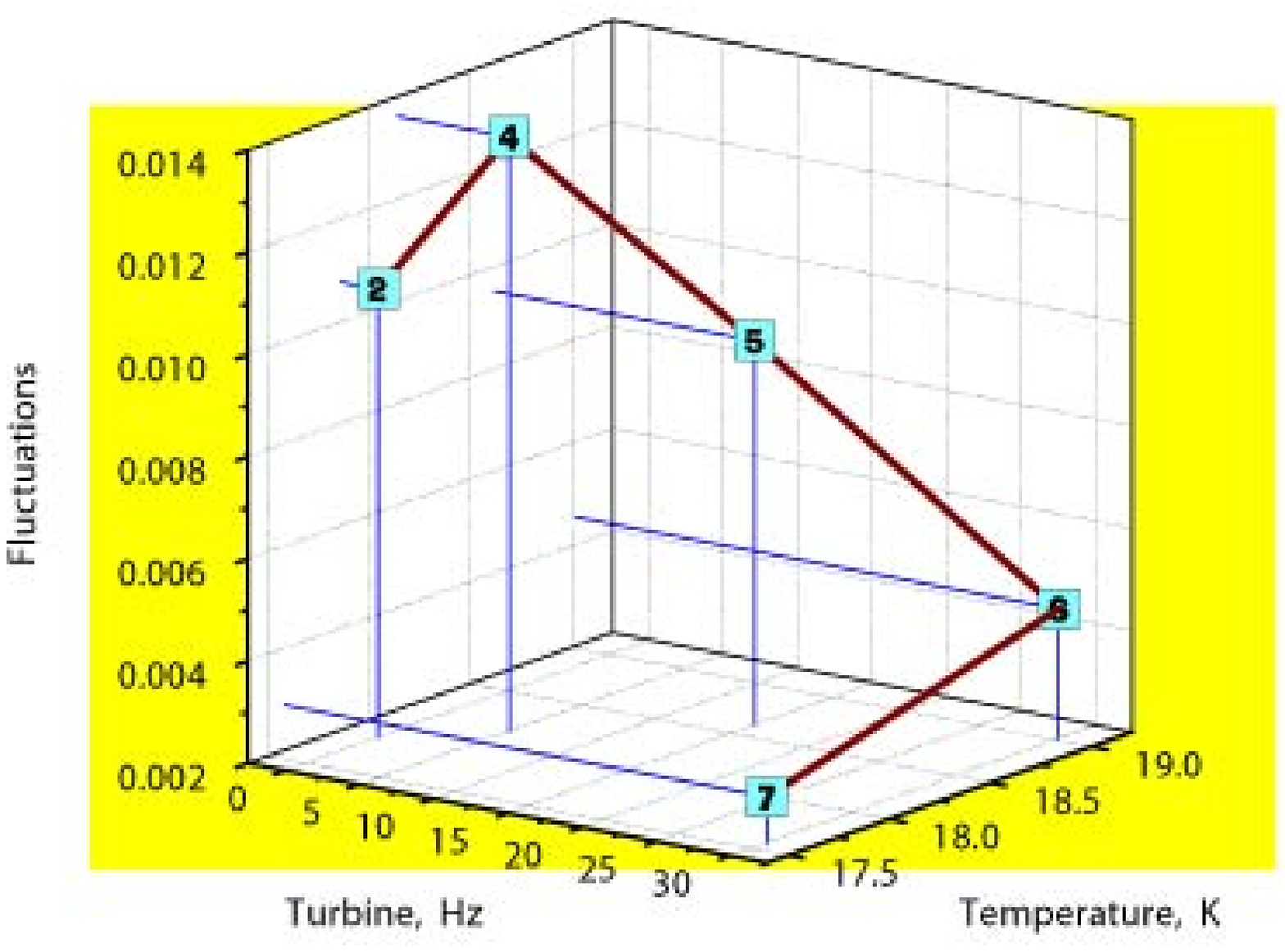}
   \end{center}
   \caption{Fluctuations at 20 $\mu$A.}
  \label{fig:flut}
\end{figure}
and low mass flow (natural convection mode) to a value of
$\sigma$~=~$\delta \mathcal{L}/\mathcal{L}$~=~0.0023 at high
hydrogen mass flow and 20~$\mu$A electron beam current
corresponding to an improvement of a factor of 14 due to the
turbulent mixing of cold and warm hydrogen in the target cell
caused by the higher mass flow. We studied the target density
fluctuations at 20~$\mu$A in more detail since the working point
of the A4 parity violation experiment at the MAMI electron
accelerator in Mainz corresponds to an electron beam current of
20~$\mu$A. The results of those measurements are shown in
Fig.~\ref{fig:flut}. As expected, the target
density fluctuations diminishes with the hydrogen mass flow
(tangential pump speed) due to higher degree of
turbulent mixing in the target cell. The observed decrease of the
fluctuations with the hydrogen temperature is also expected since
the temperature difference of the hydrogen to the boiling point
temperature is larger corresponding to more subcooled hydrogen.
\subsubsection{Influence of Electron Beam Cross Section and Exact Beam Position}
\begin{figure}[htbp]
  \begin{center}
    \includegraphics[width=0.8\textwidth]{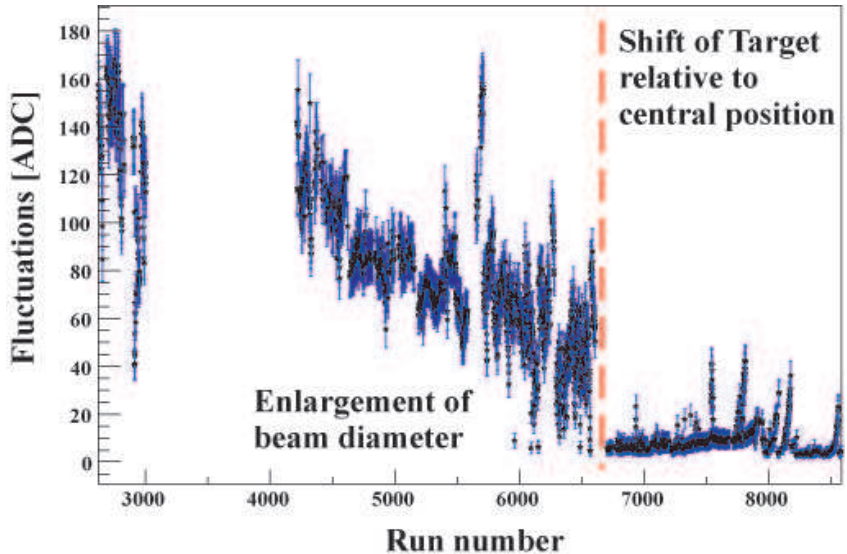}
    \caption{Investigations of target density fluctuations.}
    \label{fig:targetboil}
  \end{center}
\end{figure}

As estimated in Sec.~\ref{chap:beamheating} the size of the cross
section of the electron beam should have an essential influence
on the target density fluctuations. Fig.~\ref{fig:targetboil}
shows the the measured absolute fluctuations over a running period
of about 1000 hours of 854.3~MeV electron beam as a function of
the run number. There are two regions in
Fig.~\ref{fig:targetboil}, which are separated by the vertical
dashed line at run 6600. For the runs up to run 6600, we tried to
widen the electron beam diameter or electron beam cross
section. The change of the electron beam diameter on the target
position
is not straightforward, since besides the beam diameter on the target
other beam parameters like divergence, energy stability, helicity correlated
position and angle fluctuations and also the beam tune in the accelerator
have to be optimized at the same time. Over the
running time of a few months, we were able to continuously increase the beam diameter
of the electron beam on the target by studying the beam tune in the
electron accelerator and in the beam optics of the focusing and bending
magnets in the beam line to our accelerator hall. The effect of continuously widening the
beam cross section can be be seen in Fig.~\ref{fig:targetboil} between run 2800 and run 6600
\begin{figure}[b]
   \begin{center}
   \includegraphics[width=0.8\textwidth]{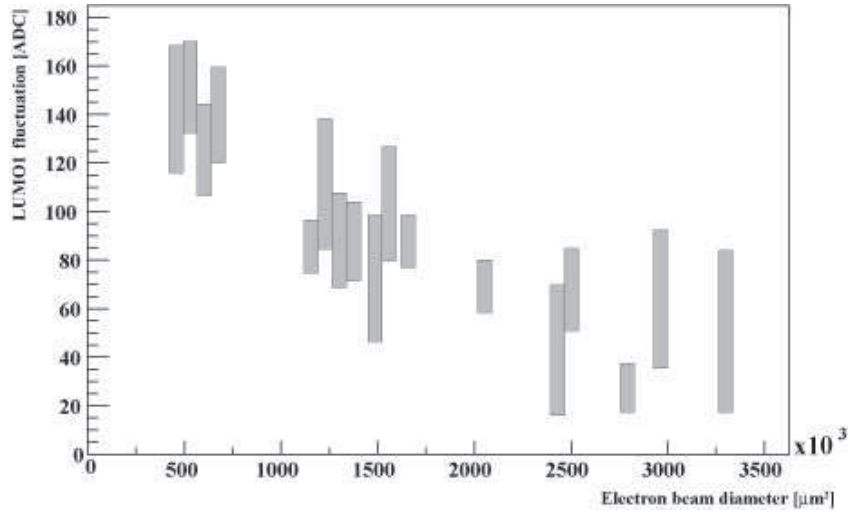}
   \end{center}
   \caption{Measured fluctuations of the hydrogen target luminosity
   as a function of the electron beam surface.}
  \label{fig:diameter}
\end{figure}
where one recognizes a steady decrease of the measured absolute fluctuations
which is caused by the increase of the beam diameter over the time
of the runs. In order to verify this, the measured absolute
fluctuations are plotted as a function of the measured area of the
elliptical beam cross section in Fig.~\ref{fig:diameter}. On can
clearly recognize a correlation between the measured fluctuations
and the size of the beam area. A further reduction of the target
density fluctuations was possible after we found, that there is a
cool spot on the entrance or exit window of the target cell. When
we move the electron beam 1~mm vertically out of the center of the
target cell, the measured fluctuations decrease by an additional
factor of 4-8 as can be seen in Fig.~\ref{fig:targetboil} after
run 6800. This can be explained tentatively by the nonsymmetric flow of the
hydrogen inside the target cell, since right at the axis of
symmetry all transversal directions of movement are equivalent to
each other. This leads to a suppression of transversal movement and
forms dead zones at windows. The asymmetric inlet in the target cell
causes a rotation of the hydrogen stream around the axis of symmetry, that is a
transversal velocity component. When the beam is moved from the axis
of symmetry, we see the effect of this rotation on
heat removal. The
\begin{figure}[btp]
   \begin{center}
   \includegraphics[width=0.8\textwidth]{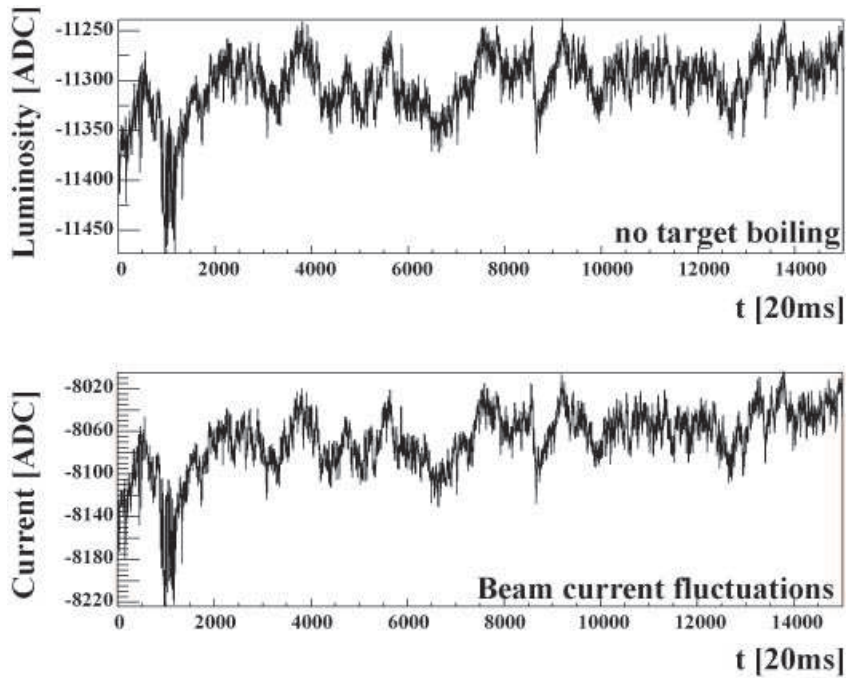}
   \end{center}
   \caption{Comparison of the signals of luminosity monitor (downstream the hydrogen target)
   and beam current monitor (upstream of hydrogen target).}
  \label{fig:complumipimo}
\end{figure}
resulting low level of residual target density fluctuations can best
be illustrated by the two plots in Fig.~\ref{fig:complumipimo}. The upper plot shows the
luminosity monitor signal of the water Cherenkov monitor at small scattering
angles downstream the hydrogen target as a function of time in units of 20~ms integration gates.
The lower plot shows the signal of the electron beam current RF-cavity monitor
in the accelerator beam line 3~m upstream the hydrogen target.
The luminosity signal follows even in details the electron beam current
signal which means that the contribution of target density fluctuations
to the observed luminosity fluctuations is very small.
A quantitative result on the luminosity fluctuations in the 20~ms of the integration time window
can be yielded from a correlation analysis of the PIMO and LUMO signals
of Fig.~\ref{fig:complumipimo}.
\begin{figure}[btp]
   \begin{center}
   \includegraphics[width=\textwidth]{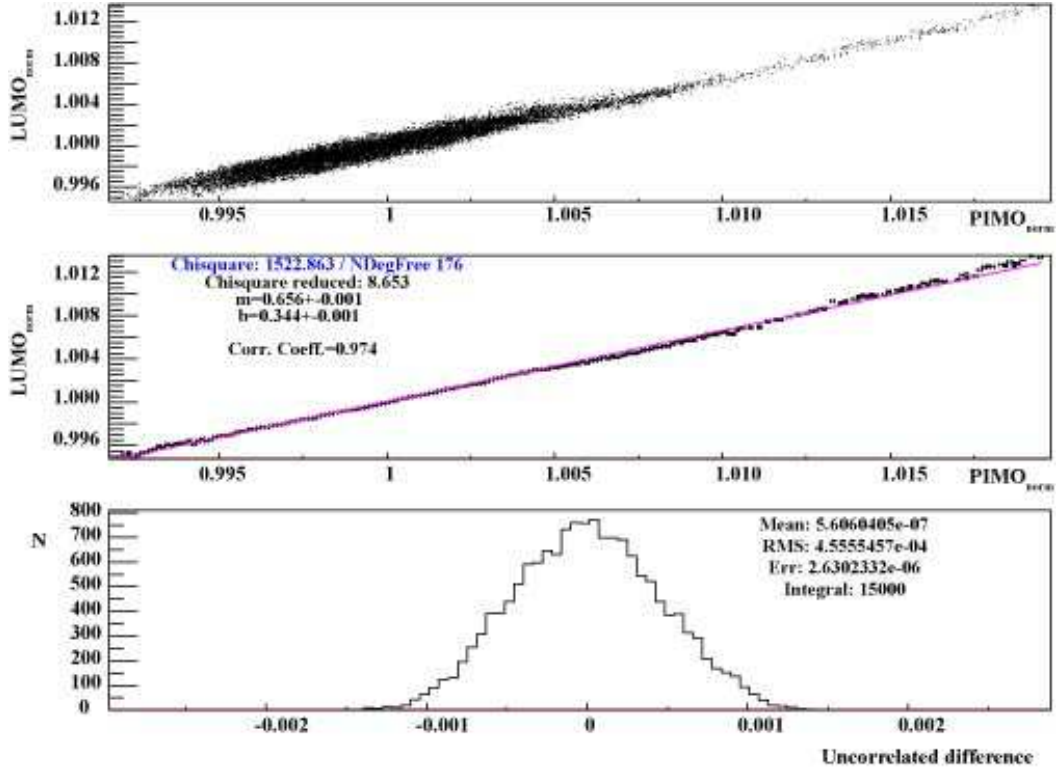}
   \end{center}
   \caption{
   Correlation analysis of the measured PIMO and LUMO data from Fig.~\ref{fig:complumipimo}.
   The upper plot shows a scatter plot. The linear correlation comes from the
   electron beam current fluctuations. For the middle plot, a straight line has been
   fitted to the averaged data of the upper plot in order to disentangle
   correlated beam current fluctuations from uncorrelated intrinsic monitor fluctuations.
   The lower plot shows the
   histogram of the differences of the data (upper plot) and the straight line (middle plot) (see text).}
  \label{fig:pimolumocorrelation}
\end{figure}
which is shown in Fig.~\ref{fig:pimolumocorrelation}.
The upper graph shows a scatter plot of the data from
Fig.~\ref{fig:complumipimo}. The luminosity monitor \#~5 is
for this plot normalized to its mean value as well as the
current monitor signal PIMO, which is normalized to its mean
value too. The linear correlation arises from electron beam current
variations during this run. The width of the band of points in the
scatter plot is a measure for the individual fluctuations
of the monitors either due to their intrinsic noise or due to
residual target density fluctuations.
A linear fit to the averaged data (middle plot) is used to
disentangle the correlation due to beam current fluctuations.
For each point in the upper plot the closest difference
of the data points to the straight line fit is calculated.
These histogrammed differences are shown in the lower plot.
The width of this histogram $\Delta PL$ is the independent (quadratic) sum
of the three contributions: a) the intrinsic noise of the current monitor PIMO ($\Delta P/P$),
b) the intrinsic noise of the luminosity monitor LUMO $\Delta L_5/L_5$, and
c) the residual target density fluctuations ($\Delta \rho/\rho$).
The observed RMS of the histogram of the normalized differences (lower plot)
in Fig.~\ref{fig:pimolumocorrelation}
is $\Delta PL = 4.56 \times 10^{-4}$.  If the intrinsic noise of both
PIMO and LUMO monitor would be zero, $\Delta PL$ would directly correspond
to the residual target density fluctuations. The intrinsic noise of the luminosity monitor
can be determined by applying the same method as in Fig.~\ref{fig:pimolumocorrelation}
to the pair of luminosity monitors \#~5 and \#~6 to $\Delta L_5/L_5 = 1.25\times 10^{-4}$.
The intrinsic PIMO noise can be determined similarly by combining PIMO \#~27 with PIMO \#~08
in the beam line further upstream to $\Delta P/P = 1.97\times 10^{-4}$.
The residual target density fluctuations within the integration time window of
20~ms can be calculated from
\begin{eqnarray}
 \Delta \rho/\rho &=& \sqrt{\Delta PL^2 - (\Delta L_5/L_5)^2 - (\Delta P/P)}
\end{eqnarray}
to $\Delta \rho/\rho = 3.92 \times 10^{-4}$ corresponding to
$\Delta \rho/\rho = 3.20\times 10^{-6}$ in the data taking run time of five minutes.

\section{Summary}
We have designed a new cryogenic liquid hydrogen target system for
the A4-Experiment at the MAMI accelerator in Mainz. The special
requirements posed by the smallness of the measured parity
violating cross section asymmetry of order 10$^{-6}$ has put
special demands on the design and construction of the target
cooling system. We have chosen the new approach which uses a high
turbulent flow in the target cell in order to enhance the
transverse mixing of the cold target liquid and thus enhancing the
transverse heat transfer. This new approach allows in
combinations with an enlarged beam spot slightly off center
on the target to abandon
rastering of the narrow beam spot on the target cell. In tests
with and without electron beam we have been able to reduce the
target density fluctuations from a factor 23 larger then the
electron beam fluctuations to negligibly lower than the electron
beam fluctuations. We plan on further parity violation
measurements including the use of deuterium and a cell with 20~cm
length at backward angles. The stability and the low level of
target density fluctuations of the liquid hydrogen target as it
has been presented in this paper has shown to fulfill the high
demands of a parity violation experiment.
\section{Acknowledgements}
This work has been supported by the Deutsche Forschungsgemeinschaft
in the framework of the SFB 201 and the SPP 1034.
We are grateful to V.~Lobashev for useful discussions and participation
in a very early state of the experiment.

\bibliographystyle{elsart-num}
\bibliography{nim_target_altarev}% Produces the bibliography via BibTeX.

\end{document}